# Turbulent Clustering of Particles and Radiation Induced Ignition of Dust Explosions

**Liberman M. [1]*, Kleeorin N[2,1], Rogachevskii I.[2,1], Haugen N-E.[3]**

[1] *Nordita, Stockholm University, Roslagstullsbacken 23, 10691, Stockholm, Sweden*
[2] *Department of Mechanical Engineering, Ben-Gurion University of the Negev, P. O. Box 653, Beer-Sheva 84105, Israel*
[3] *SINTEF Energy Research, N-7034 Trondheim, Norway*
*Corresponding author email: mliber@nordita.org*

**ABSTRACT**

Since detonation is the only established theory that allows rapid burning producing a high pressure that can be sustained in open areas, the generally accepted opinion was that the mechanism explaining the high rate of combustion in dust explosions is deflagration-to-detonation transition. In the present work we propose a theoretical substantiation of an alternative mechanism explaining the origin of the secondary explosion producing high speeds of combustion and high overpressures in unconfined dust explosions. We show that the clustering of dust particles in a turbulent flow ahead of the advancing flame front gives rise to a significant increase of the thermal radiation absorption length. This effect ensures that clusters of dust particles are exposed to and heated by radiation from hot combustion products of the primary ignited flame for a sufficiently long time to become multi-point ignition kernels in a large volume ahead of the advancing flame. The ignition times of a fuel–air mixture caused by radiatively heated clusters of particles is considerably reduced compared with the ignition time caused by an isolated particle. Radiation-induced multipoint ignitions of a large volume of fuel–air ahead of the primary flame efficiently increase the total flame area, giving rise to the secondary explosion, which results in the high rates of combustion and overpressures required to account for the observed level of overpressures and damage in unconfined dust explosions, such as for example the 2005 Buncefield explosion and several vapour cloud explosions of severity similar to that of the Buncefield incident.

## INTRODUCTION

Dust explosions occur when an accidentally ignited flame propagates through a cloud of fine particles suspended in gaseous fuel-air mixtures [1-3]. Dust explosions have been significant hazards for centuries in the mining industry and in grain elevators. Currently the danger of dust explosions is a permanent threat in all those industries in which powders of fine particles are involved. Despite intense investigations over more than 100 years the mechanism of dust explosions still remains one of the main unresolved problem. It is known that unconfined dust explosions consist of a relatively weak primary explosion followed by much more severe secondary explosions. While the hazardous effect of the primary explosion is relatively small, the secondary explosions may propagate with a speed of up to 1000 m/s producing overpressures of over 10 atm, which is comparable to the pressures produced by a detonation. However, analysis of the level of damages indicates that a detonation is not involved, while normal deflagrations are not capable of producing such high velocities and observed overpressures and damages. The possibility that DDT occurs in large explosions in industrial accidents has generally only been recognised for highly reactive fuels such as hydrogen and ethylene, and models have been validated against a range of experimental data obtained from laboratory-scale experiments. Applying such data to industrial accidents is to take the models beyond their validation range, where they cannot be used as predictive tools. In the particular case of the 2005 Buncefield fuel storage explosion investigators and forensics teams were able to collect a large amount of data and evidences, providing a unique valuable information about the timings and damage data in the event [4, 5]. A detailed analysis of physical damage and data available from CCTV cameras led to conclusions [5,6] that scenarios based on detonation [7,8,9] and the type of the observed damage are not consistent with what would occur in a detonation, and that 'the combustion in Buncefield was unsteady (episodic), with periods of rapid flame advance being punctuated by pauses.

In this paper we consider an alternative mechanism of dust explosions due clustering of dust particles in a turbulent flow ahead of the advancing flame. The turbulent clustering of particles results in a significant increase of the thermal radiation absorption length. Therefore, clusters of dust particles are exposed to and heated by radiation from hot combustion products of the primary ignited flame for a sufficiently long time to become multi-point ignition kernels in a large volume ahead of the advancing flame. Radiation-induced multipoint ignitions of a large volume of fuel–air ahead of the primary flame explain the origin of the secondary explosion with periods of rapid flame advance producing high local overpressures being punctuated by pauses.

In normal practice emissivity of combustion products and radiation absorption in a fresh unburnt gaseous mixture are small and do not influence the flame propagation. The situation is different for flames propagating through a cloud of fine particles suspended in a gas mixture. Thermal radiation emitted from the flame, propagating in a particle-laden fuel/air mixture, is absorbed and re-emitted by the particles ahead of the flame with heat being transferred from the particles to the surrounding gas by thermal conduction. It was shown [10,40] that the radiative preheating may result in a considerably strong increase in both the temperature ahead of the flame and the flame velocity. For evenly dispersed particles, the maximum increase in the temperature of the gas mixture immediately ahead of the flame front can be estimated as [39] $\Delta T \approx \left((1-e^{-1})\sigma T_b^4\right) / \left[U_f(\rho_p c_p + \rho_g c_{v,g})\right]$ where $U_f$ is the normal laminar flame velocity, $\sigma T_b^4$ is the blackbody radiative flux, $\rho_g$ is the mass density of the gaseous mixture, and $c_p$ and $c_v$ are the specific heats of particles and the gas phase, respectively.

For the evenly dispersed particles the intensity of the radiant flux decreases exponentially on the scale of the order of the radiation absorption length, $L_a = 1/\langle\kappa\rangle \approx (2\rho_p/3\rho_d)d_p$, where $\langle\kappa\rangle = \sigma_p \langle N\rangle$ is the mean particle absorption coefficient, $\sigma_p \approx \pi d_p^2/4$ is the absorption cross section of the particle, $d_p$ is the particle size (diameter), $N_p \equiv \langle N\rangle$ is the mean particle number density, $\rho_p$ is the material density of the dust particles, and $\rho_d$ is the spatial particle mass density, and $\langle N\rangle$ is the mean number density of dispersed particles. The radiation absorption length for typical for dust explosions dust cloud mass densities is in the range of (1-10) cm for the evenly dispered microns size particles. Therefore, in case of the evenly dispersed particles the radiation-induced ignition of the fuel-air by the radiatively heated particles is not possible. The situation is completely different when particles are non-uniformly dispersed, for example, in the form of the optically thick dust layer separated from ahead of the flame front by the gaseous gap with sufficiently small concentration of particles in the gap. In this case the gap between the flame and the layer is transparent for radiation, and particles in the layer can be exposed to and heated by the radiation from the flame sufficiently long time to become ignition kernels. A strong explosion ahead of the advancing flame producing a high overpressure and shock waves can occur as a result of ignition of a large volume of fuel-air mixture by the ignition kernels of particles heated by radiation from hot combustion products. However, this is possible only if the penetration length of radiation, $L_{eff}$, becomes so large, that the particles are sufficiently heated by the radiation in a large volume even far ahead of the advancing flame, so that sound waves have no time to equilize pressure, $\tau_{ign} \ll L_{eff}/a_s$, where $a_s$ is the speed of sound in the flow ahead of the flame. We will show that clustering of particles in a turbulent flow ahead of the advancing flame gives rise to a strong (up to 2–3 orders of magnitude) increase in the radiation penetration length.

## TURBULENT CLUSTERING OF PARTICLES; RADIATION HEAT TRANSFER

### The mean radiation intensity

In turbulent flows ahead of the primary flame, dust particles with material density much larger than the gas density assemble in small clusters with sizes about several Kolmogorov viscous scales. The turbulent eddies, acting as small centrifuges, push the particles to the regions between the eddies, where the pressure fluctuations are maximum and the vorticity intensity is minimum. Therefore, suspended small particles in a turbulent flow tend to assemble in clusters with much higher particlenumber densities than the mean particle number density. This effect has been investigated in a number of analytical, numerical, and experimental studies [12-17].

The equation for the intensity of radiation in the two-phase flow reads (see, e.g., [61, 62]):

$$(\hat{\mathbf{s}} \cdot \nabla) I(\hat{\mathbf{s}}, \mathbf{r}) = -\left(\kappa_g(\mathbf{r}) + \kappa_p(\mathbf{r}) + \kappa_s(\mathbf{r})\right) I(\hat{\mathbf{s}}, \mathbf{r}) + \kappa_g I_{b,g} + \kappa_p I_{b,p} + \frac{\kappa_s}{4\pi} \int_\Omega \Phi(\mathbf{r}, \hat{\mathbf{s}}; \hat{\mathbf{s}}') I(\hat{\mathbf{s}}', \mathbf{r}) d\Omega', \qquad (1)$$

where $\kappa_g(\mathbf{r})$ and $\kappa_p(\mathbf{r})$ are the absorption coefficients for the gas and particles, $\kappa_s(\mathbf{r})$ is the particle scattering coefficient, $\Phi(\mathbf{r}, \hat{\mathbf{s}}; \hat{\mathbf{s}}')$ is the scattering phase function, $I_{b,g}(\mathbf{r})$ and $I_{b,p}(\mathbf{r})$ are the black body radiation intensities for gas and particles, $\hat{\mathbf{s}} = \mathbf{k}/k$ is the unit vector in the direction of radiation. Taking into account that the scattering and absorption cross sections for gases at normal conditions are very small, the contribution from the gas phase is negligible in comparison with that of particles, Eq. (1) is reduced to

$$(\hat{\mathbf{s}} \cdot \nabla) I(\hat{\mathbf{s}}, \mathbf{r}) = \kappa(\mathbf{r}) \left(I_b(\mathbf{r}) - I(\hat{\mathbf{s}}, \mathbf{r})\right), \qquad (2)$$

where $\kappa \equiv \kappa_p(\mathbf{r})$ and $I_b \equiv I_{p,b}$ depend on the local temperature.

In the mean-field approach all quantities are decomposed into the mean and fluctuating parts: $I = \langle I \rangle + I'$, $I_b = \langle I_b \rangle + I_b'$, $\kappa = \langle \kappa \rangle + \kappa'$. The particle absorption coefficient is $\kappa = \sigma_a n$, and the fluctuations of the absorption coefficient are $\kappa' = n'\sigma_a = n'\langle \kappa \rangle / \langle N \rangle$. Averaging Eq. (2) over the ensemble of the particle number density fluctuations, we obtain the equation for the mean irradiation intensity $\langle I(\hat{\mathbf{s}}, \mathbf{r}) \rangle$:

$$(\hat{\mathbf{s}} \cdot \nabla) \langle I(\hat{\mathbf{s}}, \mathbf{r}) \rangle = -\langle \kappa \rangle \left(\langle I \rangle - \langle I_b \rangle\right) - \langle \kappa' I' \rangle + \langle \kappa' I_b' \rangle. \qquad (3)$$

Subtracting Eq. (3) from Eq. (2), we obtain the equation for fluctuations $I'$:

$$(\hat{\mathbf{s}} \cdot \nabla + \langle \kappa \rangle + \kappa') I'(\mathbf{r}, \hat{\mathbf{s}}) = I_{source}, \qquad (4)$$

where

$$I_{source} = -\kappa' \left(\langle I \rangle - \langle I_b \rangle\right) + \langle \kappa' I' \rangle + (\langle \kappa \rangle + \kappa') I_b' - \langle \kappa' I_b' \rangle. \qquad (5)$$

The solution of Eq. (5) is

$$I'(\mathbf{r}, \mathbf{s}) = \int_{-\infty}^{+\infty} I_{source} \exp\left(-\left|\int_{s'}^{s} [\langle \kappa \rangle + \kappa'(s'')] ds''\right|\right) ds'. \qquad (6)$$

Expanding the exponent in Eq.(6), multiplying the obtained equation by $\kappa'$ and averaging over the ensemble of fluctuations, we obtain for the one-point correlation function $\langle \kappa' I' \rangle$

$$\langle \kappa' I' \rangle = \left[ 1 + \int_{-\infty}^{+\infty} \left( \int_{s'}^{s} \kappa'(s) \kappa'(s'') ds'' \right) \exp(-\langle \kappa \rangle |s - s'|) ds' \right] =$$
$$= -(\langle I \rangle - \langle I_b \rangle) \int_{-\infty}^{+\infty} \langle \kappa'(s) \kappa'(s') \rangle \exp(-\langle \kappa \rangle |s - s'|) ds' \qquad (7)$$

Substituting the correlation function $\langle \kappa' I' \rangle$ into Eq. (3), we obtain an equation for the mean radiation intensity

$$(\hat{\mathbf{s}} \cdot \nabla) \langle I(\hat{\mathbf{s}}, \mathbf{r}) \rangle = -\kappa_{eff} (\langle I \rangle - \langle I_b \rangle), \qquad (8)$$

where $\kappa_{eff}$ is the effective absorption coefficient, which takes into account the particle clustering in a temperature stratified turbulence

$$\kappa_{eff} = \langle \kappa \rangle \left( 1 - \frac{2\beta J_1}{1 + 2\beta J_2} \right). \qquad (9)$$

The integrals $J_1$ and $J_2$ in Eq. (9) are given by integrals of the two-point correlation function of the particle number density fluctuations $\Phi(t, \mathbf{R}) = \langle n'(t, \mathbf{x}) n'(t, \mathbf{x} + \mathbf{R}) \rangle$ [15, 18].

**Effect of particle clustering on the effective penetration length of radiation**

The two-point correlation function that accounts for particle clustering in temperature stratified turbulence was derived in [15]:

$$\Phi(R) = (n_{cl} / \langle N \rangle)^2, \text{ for } 0 \le R \le \ell_D; \ \Phi(R) = \left( \frac{n_{cl}}{\langle N \rangle} \right)^2 \left( \frac{R}{\ell_D} \right)^{-\mu} \cos \left( \alpha \ln \frac{R}{\ell_D} \right) \text{ for } \ell_D \le R < \infty, \qquad (10)$$

where $R = \mathbf{R} \cdot \hat{\mathbf{s}}$, $\ell_D = a \ell_\eta / Sc^{1/2}$ is the size of a cluster, $\ell_\eta = \ell_0 / \mathrm{Re}^{3/4}$ is the Kolmogorov turbulent scale, $Sc = \nu / D$ is the Schmidt number, and $\mu$ is expressed as a combination of the degree of compressibility of the turbulent diffusion tensor and the degree of compressibility of the particle velocity field. It was shown [15] that the maximum number density of particles attained inside the cluster is

$$\frac{n_{cl}}{\langle N \rangle} = \left( 1 + \frac{e\mu}{\pi} Sc^{\mu/2} \ln Sc \right)^{1/2}. \qquad (11)$$

Calculating integrals $J_1$ and $J_2$ in Eq. (9) using Eqs. (10, 11) we obtain the effective penetration length of radiation $L_{eff} \equiv 1 / \kappa_{eff}$

$$L_{eff} / L_a = 1 + \frac{2a}{Sc^{1/2}} \left( \frac{n_{cl}}{\langle N \rangle} \right)^2 \left( \frac{\ell_\eta}{L_a} \right) \left( 1 + \frac{\mu - 1}{(\mu - 1)^2 + \alpha^2} \right). \qquad (12)$$

where $L_a$ is the radiation absorption length for evenly despersed particles.

There are two physical effects that affect the radiation transfer: (i) transparent for radiation windows are formed between particle clusters (ii) particles inside optically thick clusters are screened from the radiation and therefore do not participate in the radiation absoption. The present theory takes into account the collective effects of turbulent clusters on radiation transfer, but do not consider the screening effect of optically thick clusters. Therefore, the obtained results give a lower limit for the increase in the penetration length of radiation, whereas the overall effect can be much stronger.

**RADIATION-INDUCED EXPLOSIONS**

In the early stage of dust explosions the combustion mode is an accidentally ignited deflagration. The pressure waves produced by the accelerating flame run ahead producing giving rise to turbulence in the flow ahead of the advancing flame. With the increase in the primary flame surface and flame velocity, the parameters of the turbulent flow ahead of the advancing flame, i.e. $u_0$, $\ell_0$, $Re$, $St$, $\nabla \langle T \rangle$ change continuously. The dust particles in the turbulent flow ahead of the flame front assemble in clusters during a time of the order of milliseconds. Figure 1 shows the ratio $L_{eff} / L_a$ versus particle size calculated for the case of isothermal turbulence for the turbulent methane-air flow at normal conditions; $\nu = 0.2 cm^2 / s$, $c_s = 450 cm / s$, $n_{cl} / \langle N \rangle = 500$, $\sigma_{T0} = 1/2$

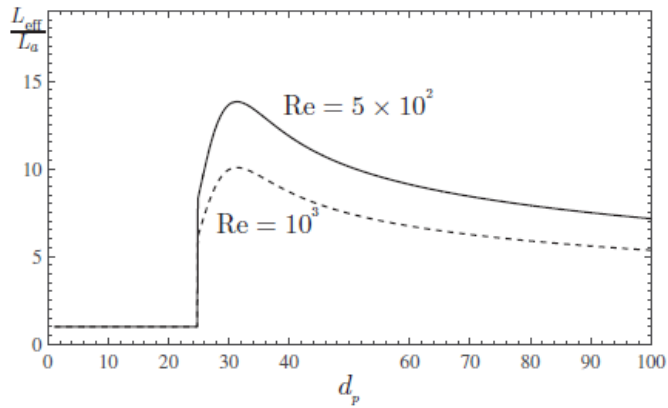

**Fig. 1.** The ratio $L_{eff} / L_a$ versus particle size for the particle clustering in isothermal turbulence with different Reynolds numbers: $5 \times 10^2$ (solid line) and $10^3$ (dashed line).

It was shown in [15] that for temperature stratified turbulence the ratio $n_{cl} / \langle N \rangle$ can increase up to three orders of magnitude. Therefore, clustering of particles in the temperature stratified turbulent flow ahead of the primary flame may increase the radiation penetration length by up to 2–3 orders of magnitude, as it is shown in Figure 2. This effect ensures that clusters of particles are exposed to and heated by the radiation from the primary flame for a sufficiently long time to become ignition kernels in a large volume ahead of the flame. The multi-point radiation-induced ignition of the surrounding fuel–air increases effectively the total flame area, so the distance, which each flame has to cover for a complete burn-out of the fuel, is substantially reduced. It results in a strong increase of the effective combustion speed, defined as the rate of reactant consumption of a given volume, and overpressures, required to account for the observed level of damages in unconfined dust explosions. If, for example, the radiation absorption length of evenly dispersed particles with spatial mass density 0.03 kg/m³ was in the range of a few centimeters, dust particles assembled in the clusters of particles, are sufficiently heated by radiation at distances up to 10–20 m ahead of the advancing flame. The experimentally measured [19] ignition time of the fuel–air by the radiatively heated particles is 100 ms for 10 μm non-reactive particles. The level of thermal radiation of hot combustion products in dust explosions, $S \sim 400 kW / m^2$ is sufficient to raise the temperature of particles by $\Delta T_p \approx 1000 K$ during $\tau_T \approx d_p \rho_p c_p \Delta T_p / S < 10 ms$ .

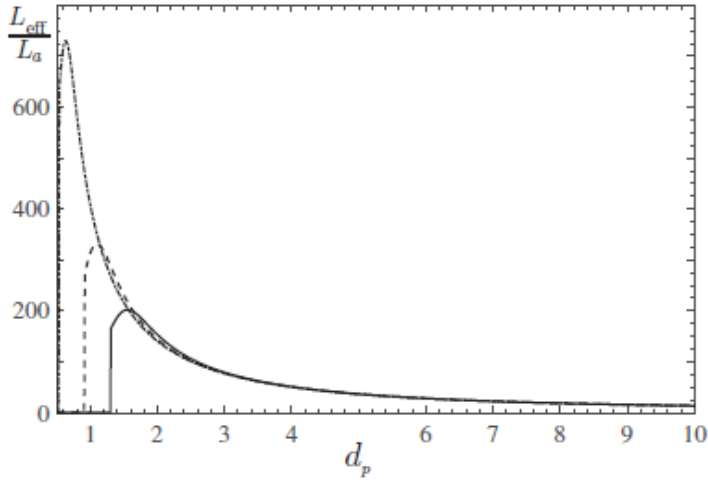

**Fig. 2.** The ratio $L_{eff}/L_a$ versus particle size for different mean temperature gradients: $|\nabla T| = 0.5$ K/m (solid), 1 K/m (dashed), 3 K/m (dashed-dotted). The particle diameter is in microns..

Figure 3 shows the dependence of $L_{eff}/L_a$ on Reynolds numbers calculated for particles of different diameter. It is seen that a significant increase of the radiation penetration length caused by particle clustering occurs within a rather narrow interval of turbulent parameters. The effect is much weaker if the flow parameters ahead of the flame front are changed and appear outside the 'range of transparency'. Such a dependence of $L_{eff}/L_a$ versus Reynolds numbers suggests possible explanation for the episodic nature of the explosion in the Buncefield incident described by Atkinson and Cusco [5].

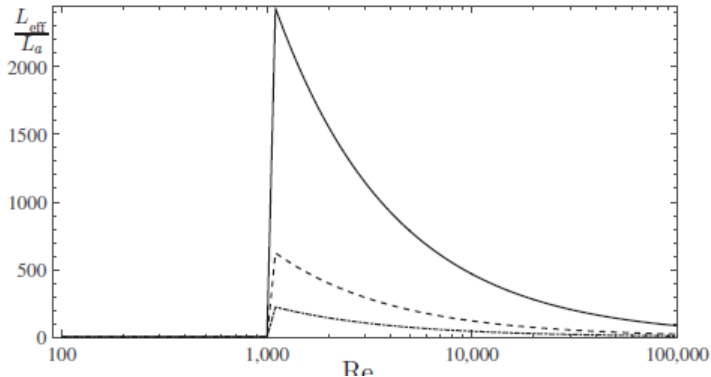

**Fig. 3.** The dependence of $L_{eff}/L_a$ on the Reynolds number for different particle diameters: 1μm (solid), 5 μm (dashed), 10 μm (dashed-dotted); the temperature gradient is 3K/m.

According to the analysis of the Buncefield explosion [5]: 'The high overpressures in the cloud and low average rate of flame advance can be reconciled if the rate of flame advance was episodic, with periods of very rapid combustion being punctuated by pauses when the flame advanced very slowly. The wide spread high overpressures were caused by the rapid phases of combustion; the low average speed of advance was caused by the pauses.' From the beginning, the parameters of the turbulent flow ahead of the advancing flame vary continuously and finally fall within the 'range of transparency' when the radiation penetration length increases considerably. Since the primary flame is a deflagration, propagating with a velocity in the order of a few meters per second, the duration of this stage is the longest timescale in the problem. During this time the particle clusters ahead of the flame are exposed to and heated by the forward radiation for a sufficiently long time to become ignition kernels in a large volume ahead of the flame initiating the secondary explosion. Since the parameters of the turbulent flow are changed after the secondary explosion, the rapid phase of combustion is interrupted until the shock waves produced by the secondary explosion dissipate. The next phase continues until the parameters of turbulence in the flow ahead of the combustion wave fall again within the interval corresponding to the 'transparent window', such that the increased ratio $L_{eff}/L_a$ caused by particle clustering provides conditions for the next secondary explosion.

## CONCLUSIONS

It is shown that the mechanism of the secondary explosion in unconfined dust explosions and large vapour cloud explosions can be explained by the turbulent clustering of dust particles. The latter include the effect of a considerable increase in the radiation penetration length, the formation of ignition kernels in the turbulent flow caused by the primary flame, and the subsequent formation of secondary explosions, which are caused by the impact of forward thermal radiation. The mechanism of multi-point radiationinduced ignitions due to the turbulent clustering of particles ensures that ignition of the gas mixture by the radiatively-heated clusters occurs rapidly and within a large volume ahead of the primary flame. The secondary explosion acts as an accelerating piston producing a strong pressure wave, which steepens into a shockwave. The intensity of the associated shock wave depends on the rate of progress of the secondary explosion and can be determined by using numerical simulations. The described scenario of unsteady combustion consisting of rapid combustion producing high overpressures, punctuated by subsequent slow combustion, is consistent with the analysis [5] of the Buncefield explosion. Although details of the real physical processes could be different, the proposed theoretical model describes the episodic nature of combustion in dust explosions observed in [5], and more importantly it captures the most important relevant physics. More detailed analysis, i.e. taking into account the particle size distribution, inter-particle collisions, possible coalescence of particles inside the clusters, and the effect of gravitational sedimentation of large particles, can only be done using numerical simulations. The obtained analytical solution can serve as a benchmark for numerical simulations of dust explosions, which do not need to rely on the simplifying assumptions.

### Acknowledgements

The authors are grateful to G. Sivashinsky, A. Brandenburg and participants of the NORDITA program 'Turbulent Combustion' for many fruitful discussions.